\documentclass{icrc2009}

\usepackage{graphicx}   
\usepackage[caption=false]{caption}    
\usepackage[font=footnotesize]{subfig} 
\usepackage{fixltx2e}
\usepackage{threeparttable}
\usepackage{url}

\newcommand{\shorttitle}[1]%
{\markboth{Proceedings of the 31\MakeLowercase{$^{st}$} ICRC, {\L}\'{o}d\'{z} 2009}{#1} }
\newcommand{\etal}{\MakeLowercase{\textit{et al. }}} 

\begin{document}
\title{Indirect Dark Matter Searches with VERITAS}

\author{\IEEEauthorblockN{R.G. Wagner\IEEEauthorrefmark{1}
			  for the VERITAS Collaboration\IEEEauthorrefmark{2}}
                            \\
\IEEEauthorblockA{\IEEEauthorrefmark{1}High Energy Physics Division,
     Argonne National Laboratory, Argonne, IL 60439-4815, USA}
\IEEEauthorblockA{\IEEEauthorrefmark{2}see R.A. Ong \etal (these proceedings) or
                  http://veritas.sao.arizona.edu/conferences/authors?icrc2009}}

\shorttitle{Wagner \etal VERITAS Dark Matter Search}
\maketitle

\begin{abstract}
A leading candidate for astrophysical dark matter (DM) is a massive particle with a mass
in the range from 50 GeV to greater than 10 TeV and an interaction cross section on the
weak scale.  The self-annihilation of such particles in astrophysical regions of high
DM density can generate stable secondary particles including very high energy gamma
rays with energies up to the DM particle mass. Dwarf spheroidal galaxies of the Local
Group are attractive targets to search for the annihilation signature of DM due to
their proximity and large DM content.  We report on gamma-ray observations taken with
the Very Energetic Radiation Imaging Telescope Array System (VERITAS) of several dwarf
galaxy targets as well as the globular cluster M5 and the local group galaxies M32
and M33.  We discuss the implications of these measurements for the parameter space of
DM particle models
\end{abstract}

 
\section{Introduction}     \label{sec:intro}

The existence of astrophysical non-baryonic dark matter (DM) has been established by its
gravitational effect on galaxy rotation \cite{rub80} and the velocity dispersions of objects
from dwarf galaxies \cite{wal07} through large galaxy clusters \cite{zwi37}.  Additional
evidence of DM existence comes from cosmic microwave background measurements
\cite{spe07} and gravitational lensing of galaxies by foreground galaxy clusters, for
example, the evident separation of dark and baryonic matter in 1E0657-558, the ``bullet
cluster'' \cite{clo06}.

At present though, the existence of dark matter is solely inferred from
its gravitational influence.  The particle nature of dark matter is yet
to be revealed through direct detection in terrestrial dark matter searches, its
production in particle accelerators, or indirect dark matter searches looking for
evidence of a flux of known particles produced by annihilation of dark matter particle
pairs.  Here we report on an indirect dark matter search carried out using VERITAS
located at the Fred Lawrence Whipple Observatory in southern
Arizona, USA, at an elevation of 1268m \cite{wee02}.  Gamma rays can be produced
either directly from the pair annhilation of weakly interacting massive particles (WIMPs)
or as secondary decay products from the primary particles produced in WIMP annihilation.
The former would produce a monoenergetic source of gamma rays equal to the WIMP mass
and would constitute definitive evidence for particle dark matter.  The
latter mechanism would produce a spectrum of gamma-ray energies with a cutoff at the
WIMP mass.

For astrophysical targets, the WIMP annihilation rate and associated gamma-ray flux are
highly uncertain due to theoretical uncertainties and limited observational constraints
on the DM halo profile.  Therefore, VERITAS has surveyed a variety of
possible sources in its dark matter search.  We present here results from observations of
three dwarf spheroidal galaxies (dSph): Draco, Ursa Minor, and Willman 1; the globular cluster M5;
and the local group galaxies M32 and M33.  The emphasis of the program has been to
target dSphs due to their relative proximity and low expected background from known
astrophysical gamma-ray sources.

VERITAS can detect and measure gamma rays in the $\sim$100 GeV - 30 TeV energy range with
an energy resolution of 15-20\%, and angular resolution of $\sim0.1^\circ$ per event.
Further technical description of VERITAS can be found in \cite{acc08}.
 
\section{Dwarf Galaxy Results}     \label{sec:dsph-results}

Dwarf spheroidal galaxies typically have stellar velocity dispersions implying a high
mass-to-light ($M/L$) ratio indicative of their dynamics being dominated by dark matter.
Observations of the Draco, Ursa Minor, and Willman 1 dSph galaxies were performed in 2007 and
2008 in three and four telescope array configurations.  Details of the observations are
summarized in Table~\ref{tbl:obs}.
\begin{table}[t]
\caption{Summary of Observation Period and Observed Time for Indirect DM Search\label{tbl:obs}}
\centering
\begin{tabular}{|l|c|c|}
\hline
Source       &  Period              &  Hours Observed   \\
\hline
Draco        &   2007 Apr-May       &    22.3           \\
Ursa Minor   &   2007 Feb-May       &    26.0           \\
Willman 1    &   2007 Dec-2008 Feb  &    13.7           \\
M5           &   2009 Feb-Mar       &    15.6           \\
M33          &   2007 Nov-2008 Feb  &    15.8           \\
M32          &   2008 Oct-2009 Jan  &    13.2           \\
\hline
\end{tabular}
\end{table}
Gamma-ray candidates are selected from stereo reconstructed events on the basis
of the summed digital pulse height from the camera phototubes, the image distance
from the center of field of view, and mean scaled width and length cuts optimized
for a signal that is 3\% that of the Crab Nebula.  As the data were acquired in
``wobble'' mode in which the telescope array was pointed at an offset of $0.5^\circ$
from the targeted source, the net signal is calculated using the ``reflected region''
background model \cite{ber07}.  Using the Li and Ma eqn.~17 method \cite{li83} to calculate
the significance, no excess above background was detected from any of the three dSphs.
Gamma-ray flux upper limits were calculated at the 95\% confidence level using the
bounded profile likelihood ratio statistic developed by Rolke~\cite{rol05}.
Table~\ref{tbl:results} summarizes the results for each of the three dwarf galaxies.
Following the formalism of \cite{woo08} we set limits in the WIMP parameter space
$(m_\chi, \langle\sigma v \rangle)$:
\begin{eqnarray*}
\lefteqn{\frac{\langle\sigma v \rangle}{3\times10^{-26}\mbox{cm}^3 \mbox{s}^{-1}} < }    \\
  &     &   R_\gamma(\mbox{95\% C.L.}) \left(\frac{m_\chi}{\mbox{100 GeV}}\right)^2      \\
  &     &   \times \left(\frac{1.45\times10^4\mbox{GeV}}{J}\right)                       \\
  &     &   \times \left\{\phi_{1\%} \int_{\mbox{200 GeV}}^\infty A(E)\left[\frac{dN(E,m_\chi)/dE}{10^{-2}\mbox{GeV}^{-1}}\right]dE\right\}^{-1},
\label{eqn:limit}
\end{eqnarray*}
where $\phi_{1\%} = 6.64\times10^{-12} \mbox{cm}^{-2} \mbox{s}^{-1}$ is 1\%
of the integral Crab Nebula flux above 100 GeV \cite{hil98}, $A(E)$ is the
energy-dependent effective collecting area, and $J$ is a dimensionless astrophysical
factor normalized to the product of the square of the critical density,
$\rho_c = 9.74\times10^{-30} \mbox{g cm}^{-3}$ and the Hubble radius, $R_H = 4.16$ Gpc.
We provide limits on the WIMP parameter space based on the assumption of a smooth NFW
profile \cite{nav97} with $J$ values given in Table~\ref{tbl:J}.  This provides conservative
estimates of the expected flux.  Significant ``boosts'' of the flux with respect to
this smooth halo assumption are possible from halo substructure that can produce
enhancements as large as a factor of 100 \cite{str07}.

\begin{figure}[t]
\includegraphics[bb= 1 1 1125 730,width=3.0in]{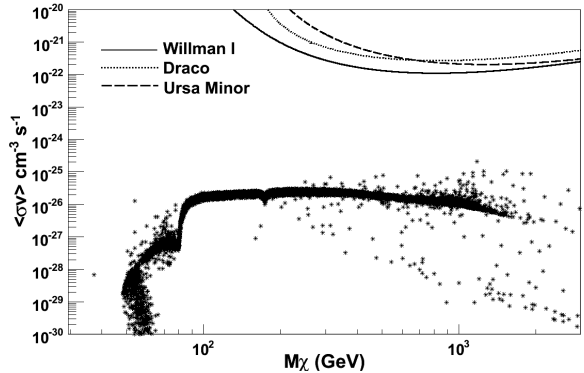}
\caption{Upper limits on $\langle\sigma v \rangle$ as a function of neutralino mass, $m_\chi$
         using a composite neutralino spectrum (see \cite{woo08} and
         the values of $J$ from Table~\ref{tbl:J}.  Black triangles represent points from MSSM
         models that fall within $\pm3$ standard deviations of the relic density measured in
         the 3 year WMAP data set \cite{spe07}. \label{fig:limit-plot}}
\end{figure}

Figure~\ref{fig:limit-plot} shows $\langle\sigma v \rangle$ limits as a function of neutralino
mass using the expression given above.  Also plotted are values from Minimal Supersymmetric
Standard Models (MSSM) generated with DarkSUSY \cite{gon04} that are consistent
with the WMAP bounds on the relic DM density \cite{spe07}.
The limits indicate that a boost factor of $\sim 1000$ would be necessary to produce a signal
within our present sensitivity.

\section{Globular Cluster and Local Group Galaxy Results}     \label{sec:other-results}

As part of the VERITAS indirect dark matter search program, we also targeted the globular
cluster, M5 and the Local Group large galaxies, M32 and M33.  Details of the exposure are
given in table~\ref{tbl:obs}.  Analysis of these data were carried out in a similar manner
to that described in section~\ref{sec:dsph-results}.  The mean scaled length and width, and
the summed pulse height selections applied to M5 are stricter than those for the dSphs,
M32, and M33.  The details of the M5 analysis are given in \cite{mcc09}.
Globular clusters and large galaxies may
have significant sources of background from standard very high energy gamma-ray producing
phenomena such as supernova remnants and compact binary objects.  Thus, while they are
worthwhile targets for an indirect DM search, interpretation of any signal would be more
complicated.  However, we found no gamma-ray excess above background of any significance
for these three targets.  The results are summarized in table~\ref{tbl:other-results}

\section{Conclusions}     \label{sec:conclu}

We have carried out a search for very high energy gamma rays from three dwarf spheroidal galaxies:
Draco, Ursa Minor, and Willman I; the globular cluster, M5; and the local galaxies M32 and M33
as part of an indirect dark matter search program on the VERITAS IACT array.
No significant excess above background was observed from any target.
We set upper limits on the flux and, for the dwarf spheroidal galaxies limits on the cross section times
velocity, $\langle\sigma v \rangle$, for neutralino pair annihilation as a function of neutralino
mass.  The $\langle\sigma v \rangle$ limits indicate that a substantial boost factor above
smooth dark halo expectations would be required of MSSM-type models.

We will continue our program in the future with an emphasis on further dSph observations.
Next generation IACT arrays now being planned such as the Advanced Gamma-ray Imaging System (AGIS)
and the Cherenkov Telescope Array (CTA) will provide an order of magnitude increase in
sensitivity over current
arrays such as VERITAS, MAGIC II, and HESS and can be expected to constrain some models of both
supersymmetric dark matter or Kaluza-Klein dark matter associated with models of universal
extra dimensions even in the conservative smooth DM halo paradigm.  Along with observations by
the Fermi Gamma-ray Space Telescope and new particle searches at the Large Hadron Collider,
prospects for understanding the nature of dark matter over the next decade look to be promising.

This research is supported by grants from the US Department of Energy, the US National Science
Foundation, and the Smithsonian Institution; by NSERC in Canada; by Science Foundation Ireland; and
by STFC in the UK.  We acknowledge the excellent work of the technical support staff at the FLWO
and the collaborating institutions in the construction and operation of the instrument.

\begin{table*}[ht]
\begin{threeparttable}
\centering
\caption{Summary Results of Dwarf Galaxy Observations\label{tbl:results}}
\begin{tabular}{|l|c|c|c|}
\hline
Quantity                                 &  Draco  &  Ursa Minor  &  Willman I    \\
\hline
Exposure (hr)                            &  18.38  &    18.91     &    13.68      \\
Signal Region (events)                   &    305  &      250     &      326      \\
Total Background (events)                &   3667  &     3084     &     3602      \\
Number Backgrd. Regions                  &     11  &       11     &       11      \\
Significance\tnote{a}                    &  -1.51  &    -1.77     &    -0.08      \\
95\% c.l. (counts)\tnote{b}              &   18.8  &     15.6     &     36.7      \\
Effective Area ($\mbox{m}^2$)            &  12518  &    16917     &    33413      \\
Energy Threshold (GeV)                   &    200  &      200     &      200      \\
Flux Limit 95\% c.l. ($\mbox{cm}^{-2} \mbox{s}^{-1}$)  & $2.82 \times 10^{-12}$  &  $1.35 \times 10^{-12}$
                               &  $2.23 \times 10^{-12}$     \\
\hline
\end{tabular}
\begin{tablenotes}
\item[a] Li and Ma eqn.~17 method \cite{li83}
\item[b] Rolke method \cite{rol05}
\end{tablenotes}
\end{threeparttable}
\end{table*}

\begin{table}[h]
\begin{threeparttable}
\caption{Parameters used for the astrophysical factor, $J$ calculation. \label{tbl:J}}
\centering
\begin{tabular}{|l|c|c|c|}
\hline
Quantity                                         &       Draco          &     Ursa Minor       &   Willman I       \\
\hline 
$R_{dSph}$ (kpc)\tnote{a}                        &        80            &       66             &     38            \\
$r_t$  (kpc)\tnote{b}                            &         7            &        7             &      7            \\
$\rho_s (\mbox{M}_\odot/\mbox{kpc}^3)$\tnote{c}  &   $4.5\times 10^7$   &   $4.5\times 10^7$   &   $4\times10^8$   \\
$r_s$ (kpc)\tnote{d}                             &      0.79            &     0.79             &    0.18           \\
$J$\tnote{e}                                     &         4            &        7             &     22            \\
\hline
\end{tabular}
\begin{tablenotes}
\item[a] Earth-dwarf galaxy distance
\item[b] The value of $J$ is negligibly changed for tidal radius, $r_t$, as low as 0.9 kpc.
\item[c] scale density
\item[d] scale radius
\item[e] $J$ is expressed as a dimensionless value normalized
                  to the critical density squared times the
                  Hubble radius, $3.832\times10^{17} \mbox{GeV}^2\mbox{cm}^{-5}$.
\end{tablenotes}
\end{threeparttable}
\end{table}

\begin{table}[h]
\begin{threeparttable}
\caption{Summary Results of Globular Cluster and Large Galaxy Observations\label{tbl:other-results}}
\centering
\begin{tabular}{|l|c|c|c|}
\hline
Quantity                                 &        M5            &      M32            &     M33           \\
\hline
Exposure (hr)                            &       15.0           &    11.29            &    11.83          \\
Signal Region (events)                   &         25           &      262            &      147          \\
Total Background (events)                &        251           &     2156            &      992          \\
Number Backgrd. Regions                  &         11           &        7            &        7          \\
Significance\tnote{a}                    &       -0.3           &     0.59            &     0.41          \\
95\% c.l. (counts)\tnote{b}              &       13.6           &     12.9            &     31.8          \\
Flux 95\% c.l. (photons $\mbox{s}^{-1}$) &  $2.5\times10^{-4}$  & $3.2\times10^{-4}$  &  $7.4\times10^{-4}$  \\
\hline
\end{tabular}
\begin{tablenotes}
\item[a] Li and Ma eqn.~17 method \cite{li83}
\item[b] Rolke method \cite{rol05}
\end{tablenotes}
\end{threeparttable}
\end{table}

\end{document}